**Extreme sensitivity of the vortex state in *a*-MoGe films to radio-frequency electromagnetic perturbation**

Surajit Dutta, Indranil Roy, Soumyajit Mandal, John Jesudasan, Vivas Bagwe and Pratap Raychaudhuri[1]

*Tata Institute of Fundamental Research, Homi Bhabha Road, Mumbai 400005, India.*

Recently, detailed real space imaging using scanning tunneling spectroscopy of the vortex lattice in a weakly pinned *a*-MoGe thin film revealed that the vortex lattice melts in two steps with temperature or magnetic field, going first from a vortex solid to a hexatic vortex fluid and then from a hexatic vortex fluid to an isotropic vortex liquid. In this paper, we show that the resistance in the hexatic vortex fluid state is extremely sensitive to radio-frequency electromagnetic perturbation. In the presence of very low-amplitude excitation above few kilohertz, the resistance increases by several orders of magnitude. On the other hand when the superconductor is well shielded from external electromagnetic radiation, the dissipation in the sample is very small and the resistance is below our detection limit.

[1] pratap@tifr.res.in

## I. Introduction

The Abrikosov vortex lattice[1,2] (VL) in a Type II superconductor behaves like soft periodic solid. In principle, thermal excitations can melt the crystalline VL into a vortex liquid with increase in temperature or magnetic field[3,4]. Yet in most superconductors observation of vortex liquid states is rare. Many experiments indicate that in bulk 3-dimensional superconductors thermal fluctuations alone cannot melt the VL into a liquid state[5,6]. Consequently, the disordered state of the VL resulting from the combined effect of thermal fluctuations and random pinning is often a vortex glass. However, in 2-dimensions the effect of thermal fluctuations can be greatly amplified. While indications of vortex liquid states have been obtained in several quasi-2D systems such as weakly pinned superconducting films[7,8] and organic superconductors[9], compelling evidence of VL melting has been obtained only in layered high-$T_c$ cuprates[10,11,12] at elevated temperatures.

Recently, evidence of vortex liquid states existing down to very low temperatures has been obtained from detailed scanning tunneling spectroscopy (STS) imaging of the vortex lattice[13] in very weakly pinned thin films of the amorphous superconductor MoGe (*a*-MoGe). Here the thickness of the film is much smaller than the bending length of the vortices such that the VL is effectively 2-dimensional. These experiments reveal that the VL in *a*-MoGe melts in two steps with increase in magnetic field or temperature. First, the hexagonal vortex solid (VS) transforms into a hexatic state with short-range positional order but quasi-long-range orientational order at fields of few kOe through proliferation of dislocations. Subsequently at a much higher field, the orientational order is lost through proliferation of disclination giving an isotropic disordered state. Furthermore, in these experiments the trajectory of each vortex has also been tracked as a function of time obtained by imaging the VL at finite time intervals. It is observed that the vortices have a finite mobility in both the hexatic and isotropic disordered vortex state, consistent with the expectation of a hexatic vortex fluid (HVF) and isotropic vortex liquid (IVL). The transitions of the VS to a HVF and HVF to IVL were also observed to leave clear imprint on the transport properties. At the first transition a small but finite linear resistance appears signaling the motion of vortices due to Lorentz force under a subcritical current drive. At the second transition, the linear resistance increases rapidly reflecting the rapid increase in the mobility of vortices as it enters the IVL state. However, despite this remarkable qualitative agreement between structural and transport measurements, a closer analysis reveals a striking discrepancy. Since STS imaging is a slow measurement, individual vortices can be imaged only when the movement of the vortices is very slow, i.e. the diffusivity ($D$) of the vortices is very small. In fact, we estimate the diffusivity, $D < 3.2 \times 10^{-21}\ m^2/s$ (see Appendix A) for fields below 85 kOe. For such a small $D$ the resultant linear resistivity ($\rho_{lin}$) is expected to be well below measurable limit. Thus in-principle, no electrical resistance should be observed in the HVF. Recently, it has been reported that the superconducting state in 2-dimensional samples of $InO_x$ and $NbSe_2$ is extremely fragile against external electromagnetic radiation[14]. When the samples were exposed to external radiation the resistance was observed to increase by several orders of magnitude which the authors attributed to overheating of the electrons. This prompted us to investigate the effect of low-frequency electromagnetic radiation on the vortex states of *a*-MoGe.

In this paper, we find that the vortex fluid states in *a*-MoGe are extremely susceptible to low frequency electromagnetic excitations. When the superconducting sample is well shielded from external electromagnetic radiations using low-pass *RC* filters, the resistance remains below the sensitivity of our measurement, till the vortices are deep in the IVL state. However when a small controlled a.c. current drive with frequency larger than few kHz is added to the d.c. measurement current the resistance in the vortex fluid states increases rapidly. We show that this sensitivity is fundamentally linked to the extremely weak pinned nature of the 2-dimensional vortex lattice and propose a model that captures the dependence of resistance on the frequency and amplitude of the a.c. excitation.

## II. Experimental Methods

*Sample.* The sample used in this study is a 20 nm thick *a*-MoGe thin film grown on surface oxidized Si substrate with $T_c$ ~ 7.2 K, similar to that used in ref. 13 . The films were grown using pulsed laser deposition using a $Mo_{70}Ge_{30}$ target prepared by arc-melting stoichiometric amounts of Mo and Ge metals. We observed that starting from arc-melted target is essential to obtain *a*-MoGe thin films with low pinning. Samples grown from commercial targets prepared by sintering MoGe powders, had similar $T_c$, but much stronger pinning. Energy dispersive X-ray compositional analysis of the thin films showed that the latter had considerable amount of dissolved oxygen which possibly gets incorporated in the sintered targets due to surface oxidation of the powder. The film used for transport measurement was capped with a 2 nm thick Si layer to prevent surface oxidation and patterned into a hall bar geometry to improve sensitivity. In contrast, the film used for STS measurements in Appendix A was directly transferred in the scanning tunneling microscope using an ultra-high vacuum suitcase without exposure to air.

*Transport measurements:* The transport measurements were performed down to 300 mK in $^3$He cryostat fitted with a superconducting solenoid of 110 kOe. The sample is mounted on a sample rotator where the angle of the normal to the film plane with respect to the magnetic field ($\theta$) can be varied from 0 – $90^0$. The default configuration for most measurements was $\theta = 0$. However, the anisotropic response of the vortex state was investigated in some measurements by varying $\theta$. To shield the sample from external electromagnetic radiation room temperature low-pass *RC* filters were installed on the electrical feedthrough connected to the sample at the entry point into the metal cryostat. The schematic of the measurement setup is shown in Figure 1. We tried 5 filters with cut-off frequencies $f_c$ varying between 340 kHz to 7.8 MHz. d.c. resistance measurements were performed using standard 4-probe method using a current source and a nanovoltmeter. The voltage drop across the voltage contacts were measured for both positive and negative polarities of current and the corresponding voltages were subtracted from one another to remove any thermopower contribution from the sample, contacts and the leads. The current was sourced either from a Keithley 6220 precision current source or from the voltage source of a Stanford Research System SR865A lock-in amplifier in series with a 22 kΩ resistor. The latter has the advantage that we can add a small a.c. voltage of a given frequency, $f$, over the d.c. voltage thereby generating a d.c. current ($I^{dc}$) superposed with a small a.c. current ( $I_f^{ac}$). Since, the combined resistance of the sample and the resistance of the connecting wire was ~ 200 Ω, this configuration resulted in a maximum inaccuracy in current of 1% as compared to a true constant current source. However, in practice the variation in current due to variation in load is much less since the primary load resistance comes from the constantan wires inside the cryostat which have very small temperature coefficient of resistance and very small magnetoresistance. By measuring standard resistors of different values we verified that the measured d.c. voltage in the nanovoltmeter is insensitive to the presence of a small a.c. voltage for frequencies > 500 Hz.

*STS measurements:* STS measurements (in Appendix A) were performed in a home-made scanning tunneling microscope[15] (STM) operating down to 350 mK and fitted with a superconducting solenoid of 90 kOe. The VL is imaged by adding a 100 μV, 2 kHz modulation voltage over the d.c. bias voltage ($V$) and recording the tunneling conductance ( $G(V) = \left.\frac{dI}{dV}\right|_V$) over the sample surface using standard lock-in technique. The d.c. bias voltage is kept close to the superconducting

coherence peaks, $V$ ~ 1.2 mV. Since the coherence peaks are suppressed inside the vortex cores, each vortex appears as a local minima in $G(V)$ in the tunneling conductance map. The d.c. tunneling current during these measurements is fixed at 50 pA. Due to stringent requirements of electromagnetic shielding to obtain reliable spectroscopic data at low bias voltages corresponding to typical superconducting energy gaps, the STM has several layers of shielding from external electromagnetic radiation in its default configuration. The entire STM is kept inside a radio-frequency shielded metal cage and all electrical inputs are fitted with r.f. filters. In addition wires leading to the sample are fitted with r.f. filters with cut-off frequency ~ 400 kHz.

## III. Results

First, we qualitatively demonstrate the extreme sensitivity of the vortex state to external electromagnetic radiation. Fig. 2(a)-(b) show the resistance versus magnetic field ( $R$-$H$ ) curves measured at 300 mK and 2 K respectively, with RC filters of different cut-off frequencies. The magnetic field, $H$, is applied perpendicular to the film plane ( $\theta = 0$ ). These measurements are performed with a d.c. current, $I^{dc} = 80\ \mu A$ which is much smaller than the flux flow critical current, $I_c$ ~ 0.5 – 1 mA (Inset Fig. 2(a)). We checked that the resistance values remain unchanged when we used lower values of $I^{dc}$ (see Appendix B). In this paper all resistance measurements are done with d.c. current of this value. The purple curve corresponds to the $R$-$H$ measured without any external filter. A finite resistance appears above 8 kOe. With increasing field the resistance increases, passes through a shallow minima and then increases again above 70 kOe. Comparing with real space STS imaging it was shown in ref. 13 that these fields corresponds to the transition from a VS to HVF state and a HVF to IVL respectively. The shallow minima in $R$ and corresponding maxima in $I_c$ are signatures of "peak effect"[16] which we discuss later. However, as we perform the measurement with different RC filters, the resistance above 8 kOe gets progressively suppressed as we use filters with lower $f_c$. For the lowest cut-off frequency, $f_c$ ~ 340 kHz, the resistance is below the sensitivity of our measurement up to 95 kOe. Thus, the resistive features associated with the structural transitions of the VL are only observed when the sample is exposed to external radio-frequency electromagnetic radiation.

To understand this effect more systematically we now investigate the response of the superconductor to a small controlled a.c. excitation. We first shield the sample using the RC filter with $f_c$ ~ 340 kHz and then add a small a.c. current, $I_f^{ac}$ (frequency $f$ < 340 kHz), to the d.c. current ( $I^{dc} = 80\ \mu A$ ). Figure 3(a)-(b) show the variation of $R$ (at 300 mK) as a function of the a.c. amplitude, $I_{100\ kHz}^{ac}$ ($f$ = 100 kHz ) in different magnetic fields. Up to 6 kOe, $R$ remains zero within the sensitivity of our measurements even in the presence of a.c. excitation. However, above this field $R$ strongly depends of on the excitation and a small $I_f^{ac}$ drives the sample from a zero resistance state to a dissipative state. Above 60 kOe, a small resistance appears even with zero excitation current, but even there the a.c excitation increases the resistance by orders of magnitude. Figure 3(c)-(d) show the variation of $R$ as a function of $f$ for a fixed amplitude, $I_f^{ac} = 6.7\ \mu A$. Again a strong frequency response is observed above 6 kOe, where the resistance increases rapidly above 10 kHz. The peak in the frequency response around 55 kHz is an artifact caused by the broad frequency response of single pole *RC* filter which slightly decreases the amplitude $I_f^{ac}$ at higher frequencies. Fig. 3(e) shows the *R-H* curves measured in the presence of different amplitudes of $I_{100\ kHz}^{ac}$ . Comparing with Fig. 2(a) one can see that increasing $I_{100\ kHz}^{ac}$ has the same qualitative effect as exposing the sample to external electromagnetic radiation. Measurements performed at 2 K (not shown here) are very similar to those performed at 300 mK. The main change with temperature is that the threshold magnetic field above which we observe the sensitivity of the superconductor to an a.c. current comes down to a lower value with increase in temperature.

In the presence of magnetic field the transport properties of the superconductor are expected to be dominated by vortex dynamics. To investigate this further, the magneto-resistive response of the sample was measured by changing the orientation of the sample with respect to the magnetic field. In Fig. 4(a) we compare the $R$-$H$ in presence of $I_{100\,kHz}^{ac} = 2.3\,\mu A$ for three orientations of the sample: (i) $H$ perpendicular to the film plane (FP) (and current); (ii) $H$ parallel to the FP and the current and (iii) $H$ parallel to the FP but perpendicular to the current. Since the thickness of our film is larger than twice the coherence length of $a$-MoGe, ξ ~ 5 nm [13], vortices can nucleate even when the magnetic field is in the plane of the film. In configuration (ii) the Lorentz force on the vortices due to the current is zero. Here the resistance remains below our sensitivity up to 100 kOe confirming that the observed resistance is caused by the flow of vortices. In configuration (iii) the Lorentz force per unit length, $\bar{J} \times \hat{n}\Phi_0$ (where $\hat{n}$ is the unit vector along $H$ and $\Phi_0 = \frac{h}{2e}$ is the flux quantum), on the vortices is the same as (i), but vortices are along the plane of the film. Here finite resistance appears only above 60 kOe. The insensitivity of the superconductor to the a.c. excitation till much larger fields than configuration (i), reflects the fact that the long vortex lines in this configuration are more strongly pinned than in the case where the magnetic field is perpendicular to the film plane. In addition, between 60 – 100 kOe the resistance shows a pronounced peak effect. Since the sample is expected to have between 1 - 2 vortices on its width, this feature is surprising and needs to be investigated further. In Figure 4(b) we plot the resistance as a function of $I_{100\,kHz}^{ac}$ at 30 kOe for different angles ($\theta$) of $H$ with the normal to the FP, but keeping the current perpendicular to $H$. The vortex length increases as $t/\cos(\theta)$, where $t$ is the film thickness. The sensitivity of the resistance to $I_{100\,kHz}^{ac}$ gradually decreases with increasing $\theta$ and goes below the sensitivity of our measurements for $\theta \geq 75^0$. These results confirm that the observed sensitivity of the superconductor is fundamentally linked to the vortex state, and the vortices in different configurations can respond differently to radiofrequency a.c. perturbation.

We would like to note that while at low fields the zero resistance state observed with $H$ perpendicular to FP is robust to small a.c. excitations, it is nevertheless possible to destroy that by increasing the amplitude of a.c. current. This is shown in Fig. 5(a) where $I_{100\ kHz}^{ac}$ is increased up to a much larger value. This raises the obvious question on whether the identification of the VL at low fields and at intermediate fields as distinct state, namely VS and HVF, is meaningful. This question is particularly important since at low fields it is difficult to quantitatively determine the metrics of positional and orientational order within the limited field of view of STS that would allow us to discriminate between a VS and HVF from structural parameters of the VL alone. To address this, in Fig. 5 (b) we plot $R$ at 300 mK in the form of an intensity plot as a function of $I_{100\ kHz}^{ac}$ and $H$. (A measurable heating on the temperature sensor placed next to the sample is only observed for $I_{100\ kHz}^{ac} > 14\ \mu A$ .) The deep blue line shows contour where $R$ exceeds 1 mΩ, the sensitivity of our measurement. We observe that threshold in $I_{100\ kHz}^{ac}$ where a measurable resistance appears rapidly decreases with magnetic field and drops below our resolution of $0.3\ \mu A$ above 17 kOe. This indicates the collapse of a pinning related energy scale which renders the vortex lattice susceptible to very small perturbations. We believe that it is this energy scale that demarcates the low field state, identified earlier as a VS with the HVF at intermediate fields. However, this needs to be confirmed through more in depth theoretical analysis.

## IV. Discussion

The most important question that arises from these measurements is why the transport properties of $a$-MoGe film displays this extreme sensitivity to very small radio-frequency perturbations, whereas bulk superconductors remain largely unaffected. In our experiments we observe that the presence of very weakly pinned 2-dimensional vortices plays a central role in the observed behavior. Since at finite frequencies vortices can dissipate energy to the electrons even with a sub-critical drive current, in principle it is possible for the sample to get overheated. A simplistic estimate of the extent of overheating required to produce an effect of the magnitude observed in our experiments can be obtained by comparing the resistance versus temperature (*R*-*T*) of the film measured, for example, at 40 kOe without filter and with an *RC* filter with $f_c$ = 340 kHz (Fig. 6(a)). Comparing the temperature at which the *R*-*T* curve with filter reaches the same resistance as that of the *R*-*T* curve without filter at 300 mK, it would require the sample to get overheated by about 4 K to observe this effect. An equilibrium heating of the sample by such large amount is impossible without manifesting as a corresponding increase in the temperature measured on the temperature sensor placed next to sample which we do not see. On the other hand, in ref. 14 it was suggested that the electrons could get overheated with respect to the phonons by several Kelvin due to weak electron-phonon coupling at low temperatures. Such a non-equilibrium overheating would not reflect on the temperature sensor which is electrically isolated from the sample. However, it will alter other superconducting properties that depend on the electronic temperature, such as the flux flow resistance, $R_{ff}$, for $I > I_c$. To verify this in Fig. 6(b) we plot the *I*-*V* characteristics in a field of 40 kOe ( 300 mK ) with and without filter. The linear slope of the *I*-*V* curve in the flux flow regime ($I > I_c$) gives $R_{ff}$ which is identical for the two curves even though the *I*-*V* characteristics at low bias (inset of Fig. 6(b)) are significantly different. From the Bardeen-Stephen model[17,18], $R_{ff}$ is given by, $R_{ff} = R_N \left(\frac{H}{H_{c2}(T)}\right)$, where $R_N$ is the normal state resistance and $H_{c2}$ is the upper critical field. Since $H_{c2}$ decreases with increase in temperature one would have expected a significant increase in $R_{ff}$ if the electrons were overheated. Similarly, the inset of Fig. 2(b) shows that the *R*- $H$ curves measured with or without filter merge at high fields, implying that $H_{c2}$ does not change significantly between the two. These observations are clearly incompatible with a large overheating of the electronic bath.

Here we propose an alternative explanation for our observations. In our experiments, we observe the extreme sensitivity of the resistance to a.c. excitation for $I << I_c$, where the Lorentz force is smaller than maximum pinning force. In this regime bundles of vortices move through thermally activated jumps[19] over the pinning barrier, $U$, such that the resistance is given by, $R_0^U = R_0 \exp\left(-\frac{U}{kT}\right)$. When a small a.c. current is added to the d.c. current, it induces small oscillatory motion of the vortex bundle inside the pinning potential well, thereby imparting it some energy, $\Delta U$. This increase in energy of the vortex bundle reduces its effective pinning to $U - \Delta U$, thus increasing the resistance to

$$R = R_0 \exp\left(-\frac{U}{kT}\right) * \exp\left(\frac{\Delta U}{kT}\right) = R_0^U * \exp\left(\frac{\Delta U}{kT}\right). \qquad (1)$$

The absolute value $\Delta U$ is difficult to calculate, since it depends on the vortex mass[20,21,22,23,24,25] ($m_v$) and the number of vortices in the vortex bundle, $N$, both of which have very high level of uncertainty. Nevertheless we can find out its dependence on the frequency, $f$, and amplitude of the a.c. current density, $J_f^{ac}$. Since vortex motion is heavily overdamped the motion of the vortex is governed by the equation[26],

$$\eta \dot{x} + kx = J_f^{ac} e^{i\omega t} \Phi_0, \qquad (2)$$

where $\omega = 2\pi f$, $t$ is time, $\eta$ is the Bardeen-Stephen viscous drag per unit length of the vortex[17], and $k$ is the restoring force per unit length due to the pinning potential (Labusch constant[27]). Solving this equation we obtain the amplitude-square of oscillation as, $x_0^2 = \left(\frac{J_f^{ac}\Phi_0}{\eta}\right)^2 \left(\frac{1}{\omega^2 + \omega_0^2}\right)$, where $\omega_0 = k/\eta$ is the depinning frequency. The energy imparted to a single vortex is $m_v\omega^2 x_0^2$. Thus for the vortex bundle,

$$\Delta U = \left(\frac{N m_v \Phi_0^2}{\eta^2}\right) J_f^{ac\,2}\left(\frac{\omega^2}{\omega^2+\omega_0^2}\right) = K J_f^{ac\,2}\left(\frac{\omega^2}{\omega^2+\omega_0^2}\right). \qquad (3)$$

In Fig. 6(c) we show $R$ as a function of $f$ for $J_f^{ac} = 1\times10^6\ A\,m^{-2}$ for $H$ = 40 kOe at 300 mK, and the corresponding fit using eqns. (1) and (3) using $R_0^U$, $K$ and $\omega_0$ as adjustable parameters. Even though the fit is not perfect, it qualitatively captures the variation of $R$ with $f$. From the fit we extract $\omega_0 = 47.2\ kHz$ which is consistent with the low depinning frequency inferred earlier from penetration depth measurements. Fig. 6(d) shows $R$ as a function of $J_{100\ kHz}^{ac}$ along with the fit with eqns (1) and (3). The fit captures the qualitative variation up to $J_{100\ kHz}^{ac} \sim 7\times10^5\ A\,m^{-2}$. The deviation at higher current densities is due to the breakdown of the harmonic approximation used in eqn. (2) caused by anharmonic terms in the pinning potential. In principle, the fits could be improved by incorporating additional refinements, for example, distribution of the size of vortex bundles which would give a corresponding distribution of $\Delta U$. However, such an exercise will add additional parameters without providing any additional physical insight. We also observe a small difference in the values of $R_0^U$ and $K$ extracted from the two fits. This is due to the fact that the 340 kHz cut-off filter does not completely eliminate the ambient radio-frequency electromagnetic radiation. Even though here the residual external radiation does not produce a detectable resistance within our measurement sensitivity, this makes the resistance of the initial state (without the application of a.c. current) somewhat arbitrary and dependent on the level of ambient noise present at the time of the measurement.

We now return to another aspect of our data namely the observation of the shallow minimum in *R*-*H* close to 72 kOe when measured without a filter (Fig. 2(a)). At the same field $I_c$ shows a maxima (inset Fig. 2(a)). A peak in $I_c$ is widely observed in weakly pinned Type II superconductors and is associated with the non-monotonic variation of collective pinning[28] close to an order-disorder transition of the VL. In $a$-MoGe, it has been shown that the peak effect in $I_c$ can be explained within collective pinning theory to arise from the HVF to IVL transition of the VL[29]. The model presented here explains why it also manifests in $R$-$H$ when additional a.c. excitation is present. Within our model the observed resistance is the TAFF resistance, $R_0^U$, exponentially amplified by the a.c. excitation. Since within the approximation of critical state model[30] the effective pinning barrier[31], $U \propto J_c B V_c$ (where $V_c$ is the Larkin volume), a maxima in the critical current results in a corresponding minima in $R_0^U$. However, this effect becomes only visible only when the resistance is amplified to measurable threshold of sensitivity due to the presence of external electromagnetic noise or controlled a.c. excitations.

Finally, it is interesting to try to identify the source of electromagnetic radiation that could give rise to the observed effect when measurements are performed without filter. We observe that using a low-pass filter with 340 kHz cut-off effectively removes the effect of ambient radiation, even though the sample responds to much lower frequencies. To understand this we first note that most wireless communicating devices, such as Wi-Fi, Bluetooth, wireless mouse and keyboard work at frequency ranges well above 100 MHz. FM radio stations which are another source of electromagnetic radiations transmit at carrier frequencies of tens of MHz. All these are easily filtered out with filters with much higher cut-off. However, in the city of Mumbai we found six short and medium wave AM radio stations transmitting in hundreds of kHz to few MHz frequency range[32]. The lowest frequency ambient radiation source that we could identify is Mumbai B (Asmita Vahini) radio channel transmitting at 558 kHz. Once the filter cut-off is set to lower than this value, the spectrum is more of less silent and any remaining radiation is not strong enough to drive the resistance above our limit of detection.

## V. Conclusion

In conclusion, we have shown that the 2-dimensional vortex states in weakly pinned $a$-MoGe thin film is extremely susceptible to external electromagnetic excitations. The most intriguing part of this study is the very low frequency and amplitude threshold of ac excitation above which we see this effect. We believe that this is due to the formation of vortex fluid states with very low mobility of the vortices. Thus even though the vortex motions at low currents in these fluid states do not produce any detectable resistance, the mobility of the vortices gets vastly enhanced in the presence of an external electromagnetic excitations. It would be interesting to explore to what extent this could be generic for other systems, such as $InO_x$ and few layered $NbSe_2$ which display similar sensitivity in the presence of magnetic field.

### Appendix A: Diffusion constant of the vortices from real space STS images

The vortex diffusion constant, $D$, which is related to the resistance observed in the mixed state of a Type II superconductor, can in-principle be obtained by tracking the motion of vortices as a function of time as was done in ref. 13. From microscopic random walk viewpoint of the Brownian motion, the Mean Square Displacement ($\langle r^2 \rangle$) is related to the Diffusion coefficient ($D$) as, $\langle r^2 \rangle = 4\, Dt$, where, $t$ is time.[33] To experimentally determine $\langle r^2 \rangle$ we capture 12 successive images of the VL using STS staying at the same field, at intervals of 15 minutes. The trajectory of each vortex as a function of time is obtained by comparing each successive image. An example, at 55 kOe, 450 mK is shown in Fig. 7.

In Fig. 8 we plot the mean square displacement normalized to the vortex lattice constant, *a,* as a function of time at different fields. At fields higher than 70 kOe, the variation is linear as expected for Brownian motion in an isotropic vortex liquid. To obtain an upper bound to the diffusion constant, we calculate the diffusion constant at 85 kOe as, $D = 3.23632 \times 10^{-21}\ m^2/s$.

We can now calculate the resistivity that would arise from the vortex motion. Assuming that the motion is thermal in origin, the mobility of the vortices is, $\mu = D/k_B T$ . This gives the effective viscosity of the vortices as, $\eta = \frac{1}{\mu d} = \frac{kT}{Dd} = 1.37063 \times 10^5\ kg\ m^{-1}\ s^{-1}$. This would give a thermally activated flux flow resistivity[18], $\rho_{TAFF} = B \times \frac{\Phi_0}{\eta} = 1.28371 \times 10^{-19}\ \Omega\ m$ , which is well below the sensitivity of our measurements. Though this kind of analysis is not completely free from our inability to determine large flux jumps given the indistinguishability of the vortices, this gives an order of magnitude estimation of the resistivity which one would expect in the vortex fluid phases. Thus given the current scenario in absence of external electromagnetic perturbation, one would expect no appearance of resistivity in the HVF phase.

**Appendix B: Resistance measurement with different d.c. currents**

To understand the effect of $I_f^{ac}$ on the resistance it is important to ensure that $I^{dc}$ is kept low enough such that non-linear resistive effects do not result from the d.c. current drive itself. In Fig. 9(a) we show *R-H* measured without filter for $I^{dc}$ = 10, 20, 50 and 80 μA. For $H < 90$ kOe, which is the range of field of interest in this work, the resistance is identical for all 4 curves within the accuracy of the measurement. This can also be seen from the linearity of the *I-V* curves shown in the inset. (For $H > 90$ kOe there is an increase in $R$ for larger $I^{dc}$ since $I_c$ comes down rapidly.) In Fig. 9(b) we show the variation of $R$ as a function of $I_{100\ kHz}^{ac}$ at 25 kOe, 300 mK using different values of $I^{dc}$. Except for the increases noise at lower currents again the curves for different $I^{dc}$ are similar. All other measurements in the paper are done with $I^{dc} = 80$ μA.

We would like to thank Vadim Geskenbein for drawing our attention to the possible discrepancy between STS and transport data. We would also like to thank Benjamin Sacepe, I Tamir and Dan Shahar and Peter Armitage for valuable discussions. Finally, we would like to thank the organizers of the workshop "The challenge of 2-dimensional superconductivity" at Lorentz Center, The Netherlands for a vibrant discussion on the effect high-frequency electromagnetic noise on the transport properties of 2-dimensional superconductors. This work was financially supported by Department of Atomic Energy, Govt of India.

SD performed the transport measurements and analyzed the data. IR performed the STS measurements and analyzed the data. JJ and VB prepared the a-MoGe film and performed basic characterization. SM carried out complimentary magnetic measurements and wrote the instrument interface codes for transport measurements. PR conceptualized the problem, supervised the project and wrote the paper with inputs from all authors.

---

[20] H. Suhl, *Inertial Mass of a Moving Fluxoid*, Phys. Rev. Lett. **14**, 226 (1965).

[21] E. M. Chudnovsky and A. B. Kuklov, *Inertial Mass of the Abrikosov Vortex,* Phys. Rev. Lett. **91,** 067004 (2003).

[22] N. B. Kopnin and V. M. Vinokur, *Dynamic Vortex Mass in Clean Fermi Superfluids and Superconductors*, Phys. Rev. Lett. **81,** 3952 (1998).

[23] M. W. Coffey, *On the inertial mass of a vortex in high-$T_c$ superconductors: closed form results*, J. Phys. A: Math. Gen. **31**, 6103 (1998)

[24] E.Simánek, *Inertial mass of a fluxon in a deformable superconductor*, Phys. Lett. A **154,** 309 (1991).

[25] D. M. Gaitonde and T. V. Ramakrishnan, *Inertial mass of a vortex in cuprate superconductors*, Phys. Rev. B **56**, 11951 (1997).

[26] J. I. Gittleman and B. Rosenblum, *Radio-Frequency Resistance in the Mixed State for Subcritical Currents,* Phys. Rev. Lett. **16**, 734 (1966).

[27] R. Labusch, Cryst. Lattice Defects **1**, 1 (1969).

[28] W. Henderson, E. Y. Andrei, M. J. Higgins and S. Bhattacharya, *Metastability and Glassy Behavior of a Driven Flux-Line Lattice,* Phys. Rev. Lett. **77**, 2077 (1996).

[29] See supplementary material in ref. 13.

[30] C. P. Bean, *Magnetization of Hard Superconductors*, Phys. Rev. Lett. **8**, 250 (1962).

[31] J. B. Ketterson and S. N. Song, Superconductivity (Cambridge University Press, Cambridge, 1999).

[32] For a list of radio stations in Mumbai, see http://worldradiomap.com/in/mumbai

[33] K. A. Dill and S. Bromberg, Molecular Driving Forces: Statistical Thermodynamics in Biology, Chemistry, Physics, and Nanoscience, Briefings in Bioinformatics **4**, 382 (2003).

## Figure Captions

**Figure 1|** (left) Schematic drawing of the 4-probe resistance measurements setup. The current is sourced from an a.c. + d.c. voltage source in series with a 22 kΩ resistor. For measurements where no a.c. current is applied a precision current source was used instead. The d.c. voltage is measured using a nanovoltmeter. RC filters are connected on the electrical feedthrough at room temperature leading to the sample. (right) Typical frequency response of the the RC filter with $f_c = 340$ kHz.

**Figure 2|** (a) d.c. resistance, $R$, as a function of magnetic field, $H$, at 300 mK measured with RC filters of different $f_c$. (*inset*) Current versus voltage (*I-V*) characteristics of the film at two different fields (thick green and red lines) from which the flux flow critical current, $I_c$, (blue curve) is calculated; $I_c$ is determined by extrapolating back the linear flux flow regime of the *I-V* curve. (b) same as (a) but at 2 K; (*inset)* the same data as in the main plot but plotted in semi-log scale.

**Figure 3|** (a)-(b) d.c. resistance, $R$, as a function of the amplitude of a.c. current, $I_{100\ kHz}^{ac}$, in different magnetic fields at 300 mK. The lines in the 3-dimensional plot, (b), correspond to the variation of resistance for fixed $I_{100\ kHz}^{ac}$ as a function magnetic field. (c)-(d) d.c. resistance as a function of the frequency, $f$, of a.c. current with a fixed amplitude of 6.7 μA, in different magnetic fields at 300 mK. The line in the 3-dimensional plot, (d), correspond to the variation of resistance for a $f = 50$ kHz as a function magnetic field. (e) *R-H* at 300 mK measured in the presence of $I_{100\ kHz}^{ac}$ of different magnitudes. Measurements are done with RC filter with $f_c = 340$ kHz.

**Figure 4|** (a) d.c. resistance, $R$, versus magnetic field at 300 mK with $I_{100\ kHz}^{ac} = 2.3\ \mu A$, in three different sample orientations with respect to the magnetic field as shown above the panel. (b) d.c. resistance as a function of the a.c. current amplitude at 30 kOe, $I_{100\ kHz}^{ac}$, with magnetic field, $H$, at different angles, $\theta$, with respect to the normal to the film plane, while the current is kept

perpendicular to the *H*. (*inset*) Schematic of the field orientation with respect to magnetic field and current; the black arrow denotes the normal to the film plane. Measurements are done with RC filter with $f_c$ = 340 kHz.

**Figure 5|** (a) d.c. resistance, *R*, as a function of $I_{100\ kHz}^{ac}$ in different magnetic fields at 300 mK. (b) Intensity plot of *R* as a function of $I_{100\ kHz}^{ac}$ and magnetic field, *H*. The blue line shows the contour for which $R \approx 1$mΩ. Measurements are done with RC filter with $f_c$ = 340 kHz.

**Figure 6|** (a) d.c. resistance, *R*, as a function of temperature, *T* measured at 40 kOe, without RC filter and with RC filter with $f_c$ = 340 kHz. The dotted line denotes the resistance measured at 300 mK when the measurement is done without a filter. (b) *I-V* characteristics of the sample at 40 kOe, 300 mK without and with the same filter; (*inset*) expanded view of the graph at low currents. (c) *R* as a function of frequency, *f*, of an a.c. current density of $J_f^{ac} \sim 1 \times 10^6$ A/m$^2$ (corresponding to $I_f^{ac} \sim 5.9$ μA) at 40 kOe, 300 mK. (*inset*) same data plotted in semi-log scale. (d) *R* as a function of $J_{100\ kHz}^{ac}$ at 40 kOe, 300 mK. The blue lines in (c) and (d) are fits using our theoretical model. In (c) and (d) we used RC filters with $f_c$ = 340 kHz.

**Figure 7|** (left panel) First image of 12 consecutive vortex images at 450 mK for 55 kOe. VL is Delaunay triangulated to find topological defects (denoted as red, green, magenta, and yellow dots) corresponding to five-, seven-, four-, and eightfold coordination. (middle panel) arrow map of the vortex motion, in which each arrow gives displacement for every vortex through individual steps of 12 consecutive images. (right panel) Enlarged view of the red box in the middle panel.

**Figure 8|** The mean square displacements in units of lattice constant, $a$, as a function of time determined from the 12 images at different magnetic fields at 450 mK. The magenta straight line is a linear fit to the 85 kOe data.

**Figure 9|** (a) Resistance, R, as a function of magnetic field 300 mK, for $I^{dc}$ =10, 20, 50 and 80 μA. The data is taken without filter. (*inset) I-V* Characteristics ( low current regime, $I << I_c$ ) at 300 mK. (b) $R$ as a function of $I^{ac}_{100kHz}$ for $I^{dc}$ = 5, 10, 20, 40 and 80 μA at 25 kOe, 300 mK.

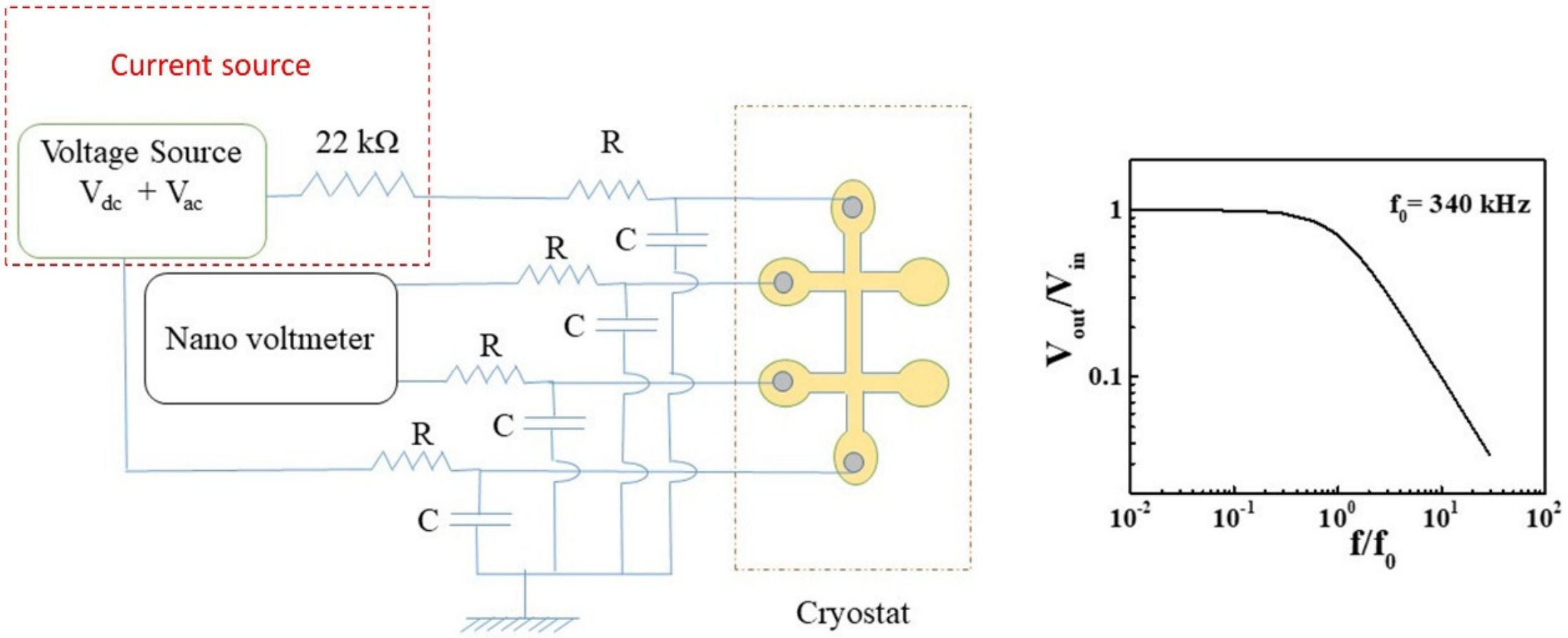

**Figure 1**

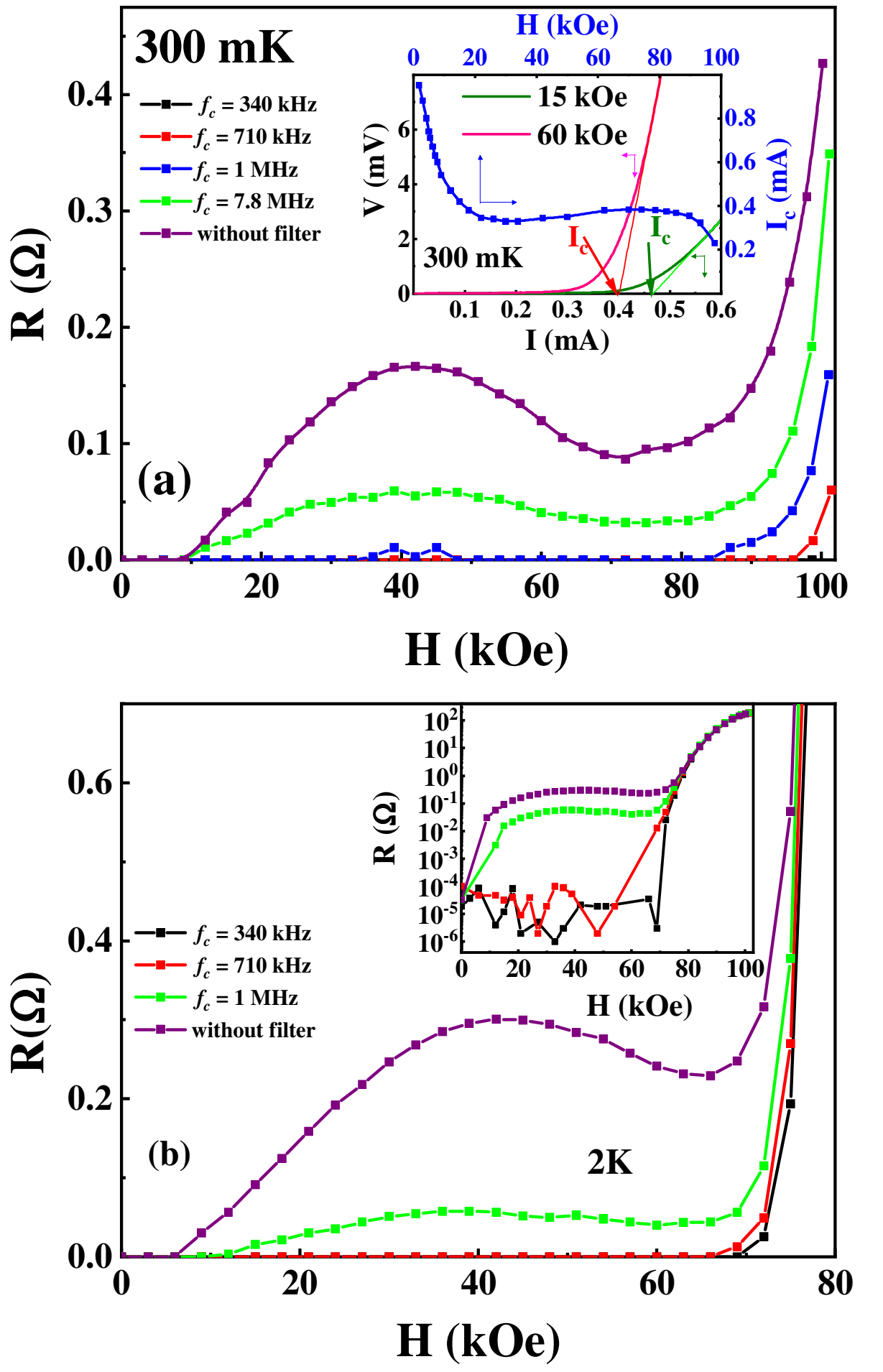

300 mK
$f_c$ = 340 kHz
$f_c$ = 710 kHz
$f_c$ = 1 MHz
$f_c$ = 7.8 MHz
without filter
H (kOe)
15 kOe
60 kOe
V (mV)
$I_c$ (mA)
300 mK
$I_c$
$I_c$
I (mA)
R (Ω)
(a)
H (kOe)
$f_c$ = 340 kHz
$f_c$ = 710 kHz
$f_c$ = 1 MHz
without filter
R (Ω)
H (kOe)
R(Ω)
(b)
2K
H (kOe)

**Figure 2**

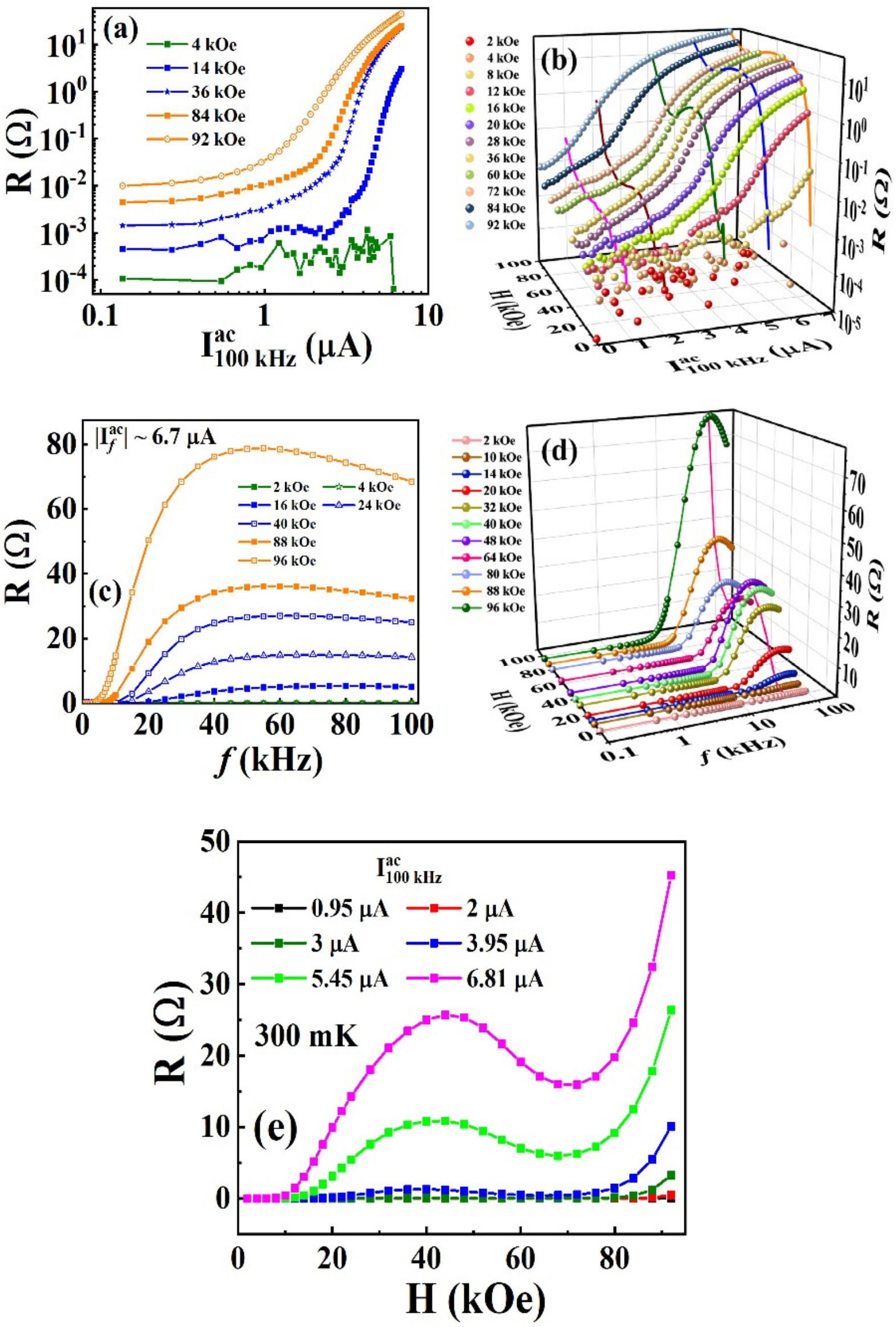

**Figure 3**

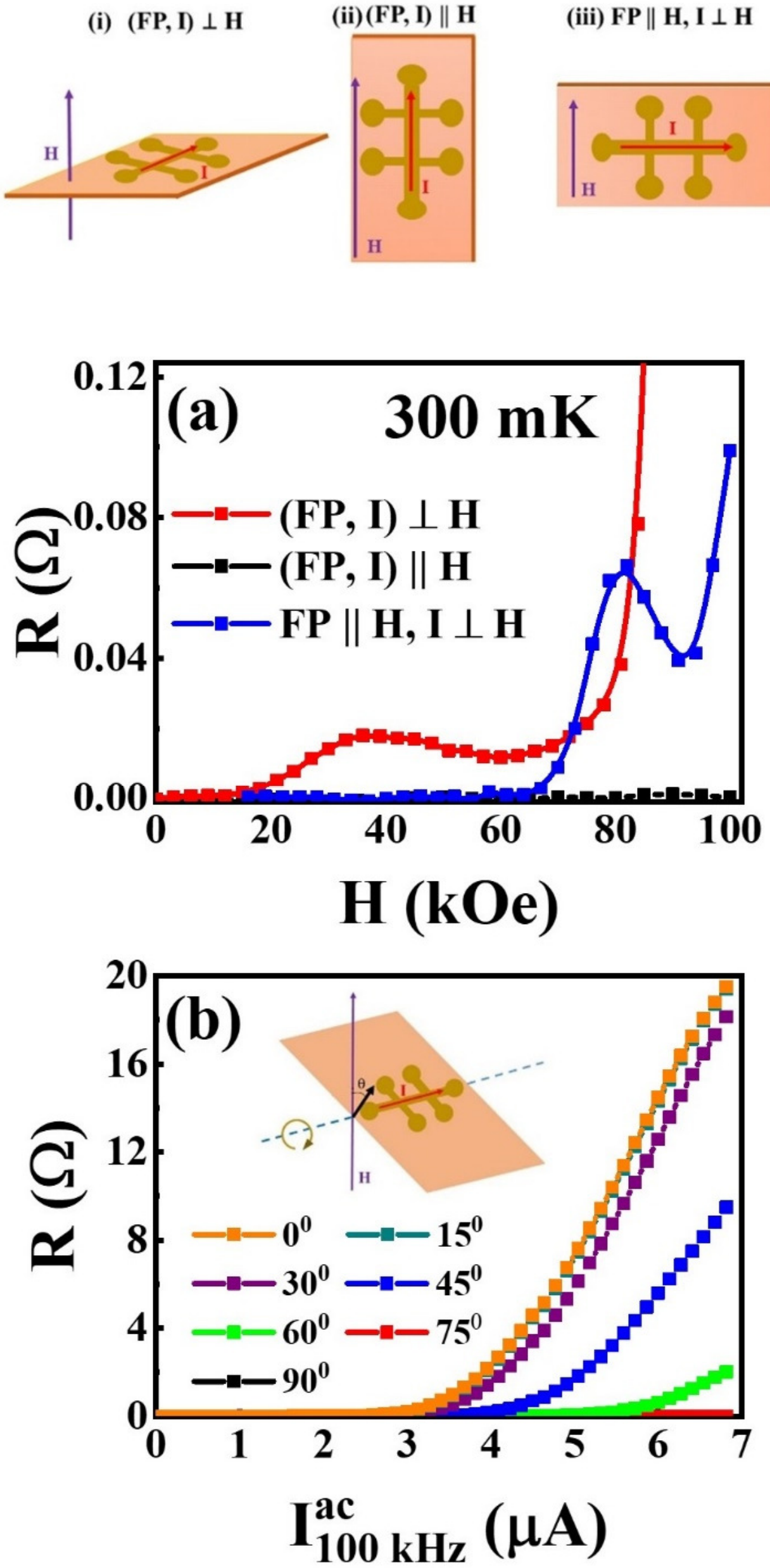

(i) (FP, I) ⊥ H
(ii) (FP, I) || H
(iii) FP || H, I ⊥ H
H
I
(a)
300 mK
(FP, I) ⊥ H
(FP, I) || H
FP || H, I ⊥ H
R (Ω)
H (kOe)
(b)
θ
0⁰
15⁰
30⁰
45⁰
60⁰
75⁰
90⁰
R (Ω)
$I^{ac}_{100\ kHz}$ (μA)

**Figure 4**

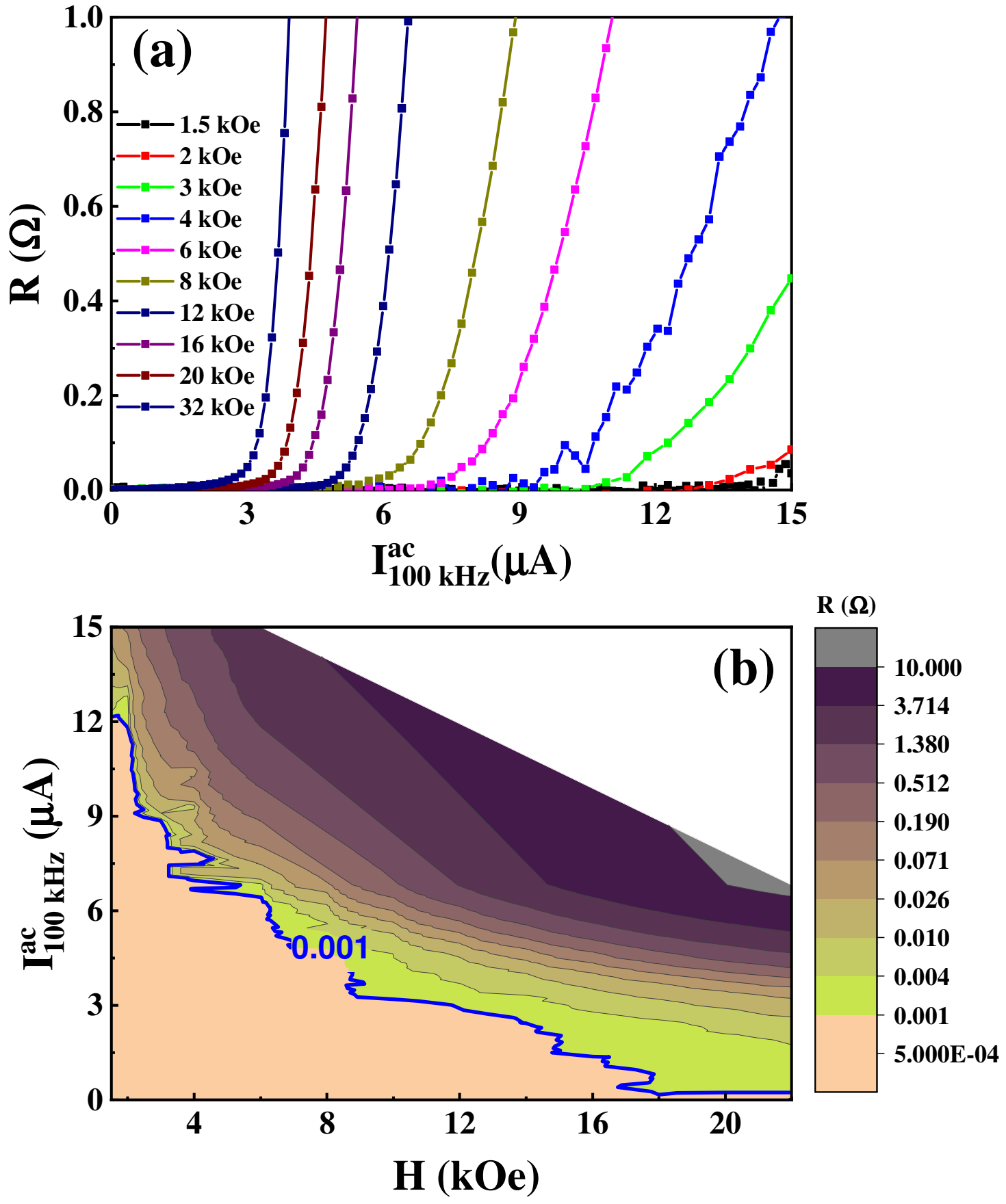

**Figure 5**

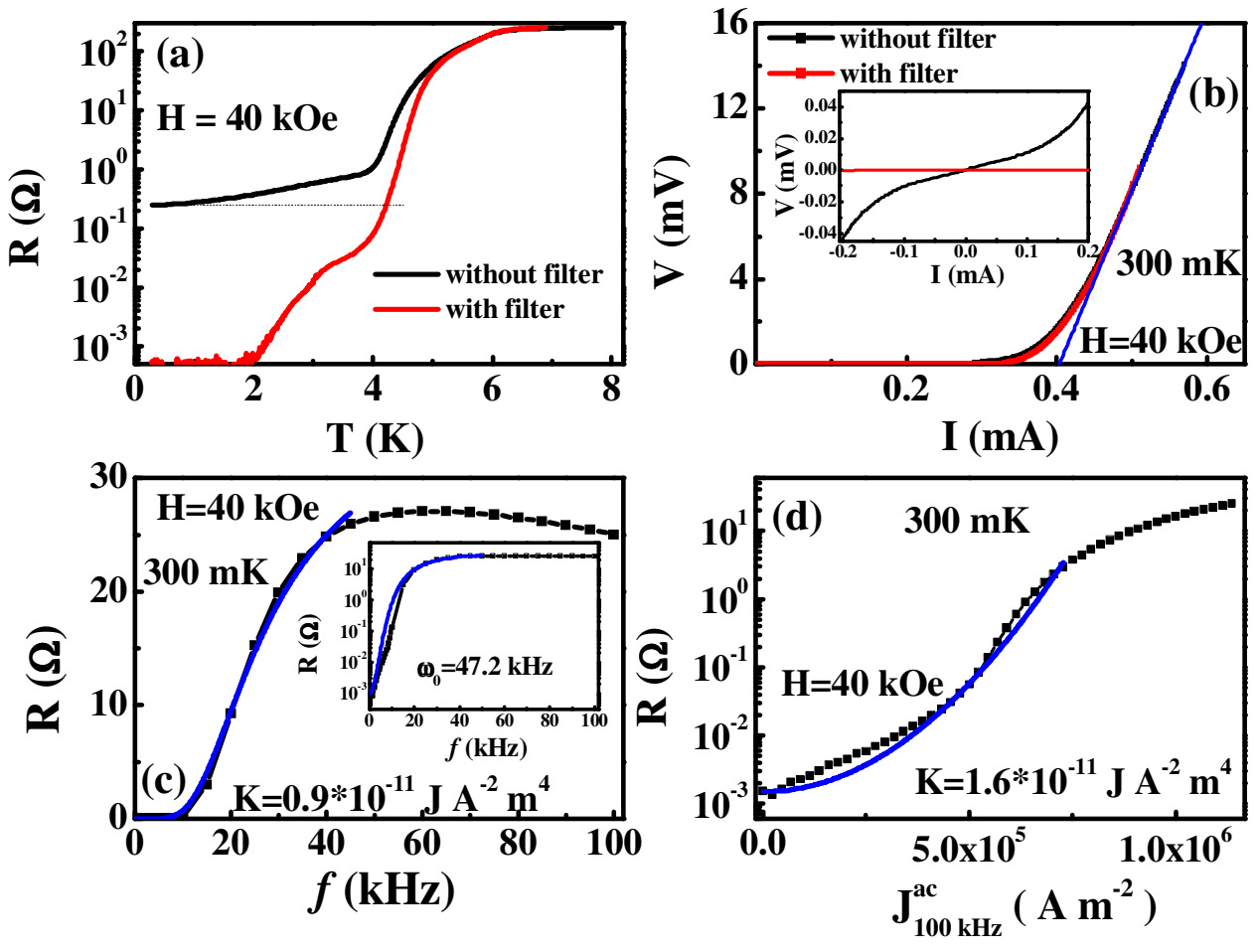

(a)
H = 40 kOe
without filter
with filter
R (Ω)
T (K)
(b)
without filter
with filter
300 mK
H=40 kOe
V (mV)
I (mA)
(c)
H=40 kOe
300 mK
ω0=47.2 kHz
K=0.9*10-11 J A-2 m4
R (Ω)
f (kHz)
(d)
300 mK
H=40 kOe
K=1.6*10-11 J A-2 m4
R (Ω)
J ac 100 kHz ( A m-2 )

**Figure 6**

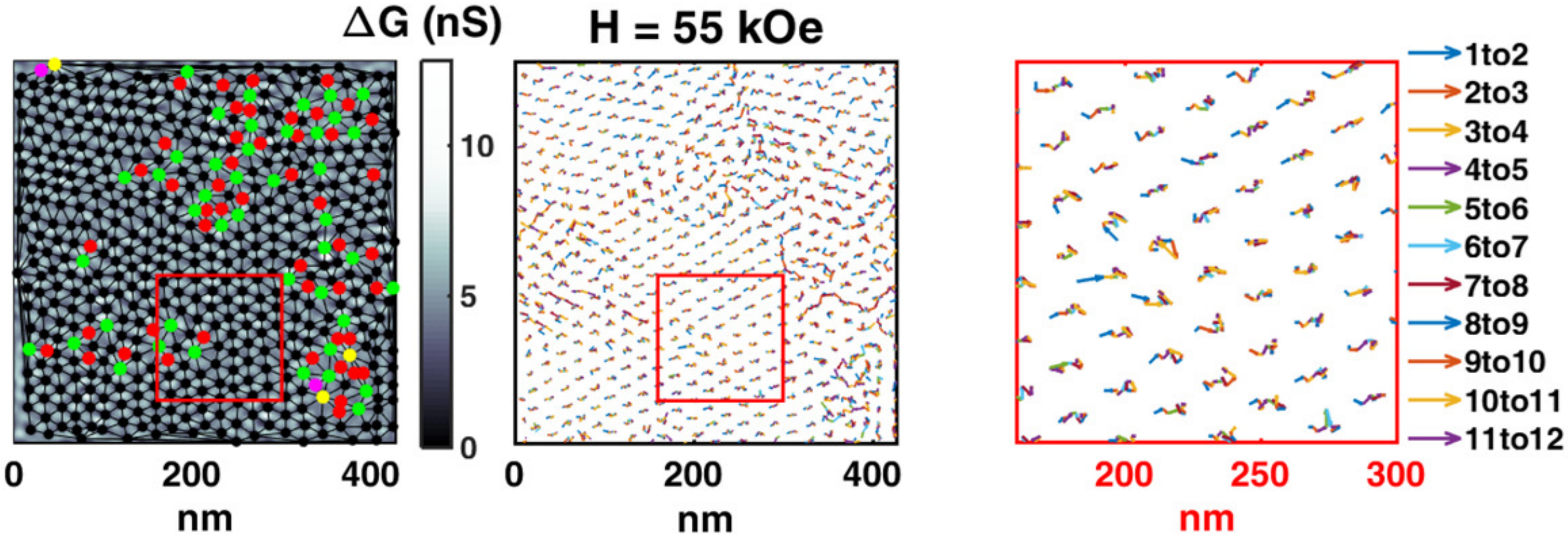

**Figure 7**

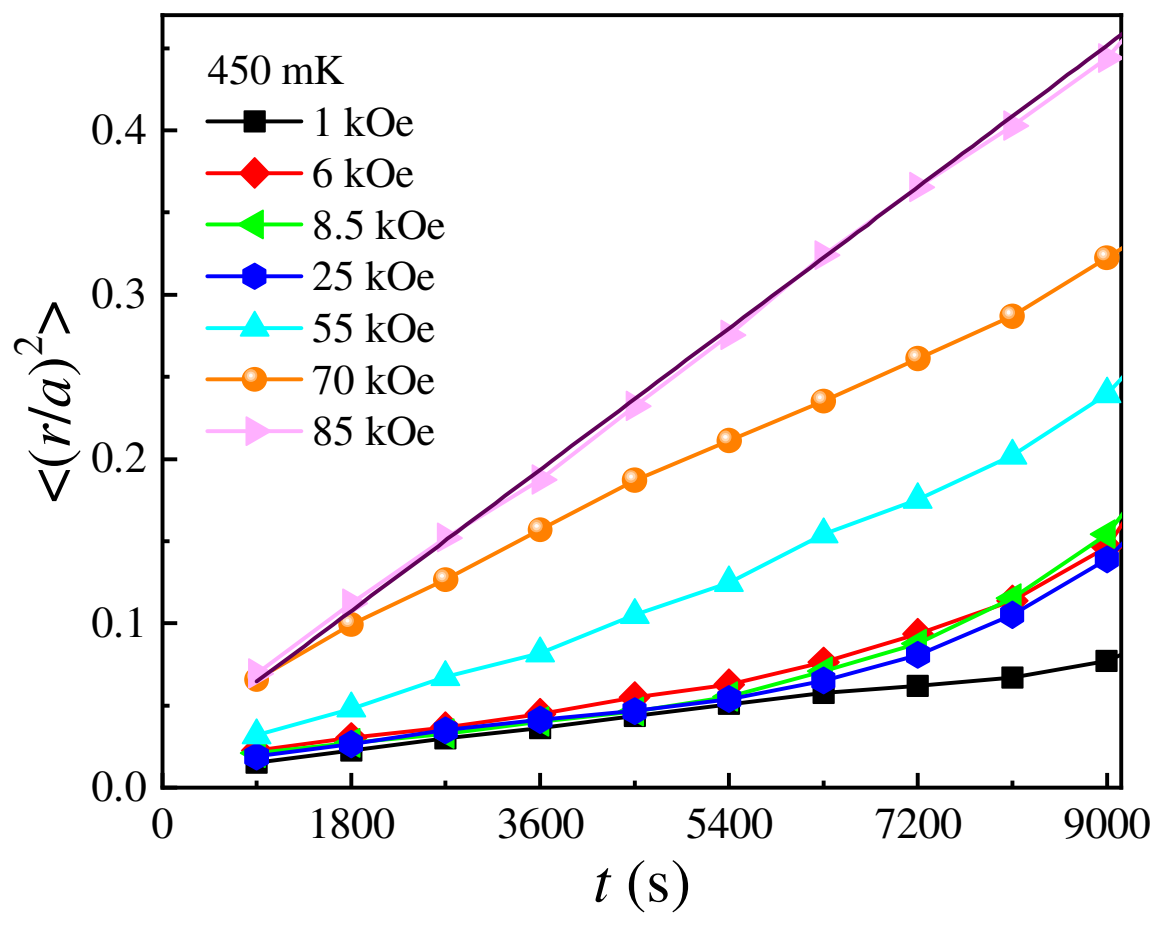

**Figure 8**

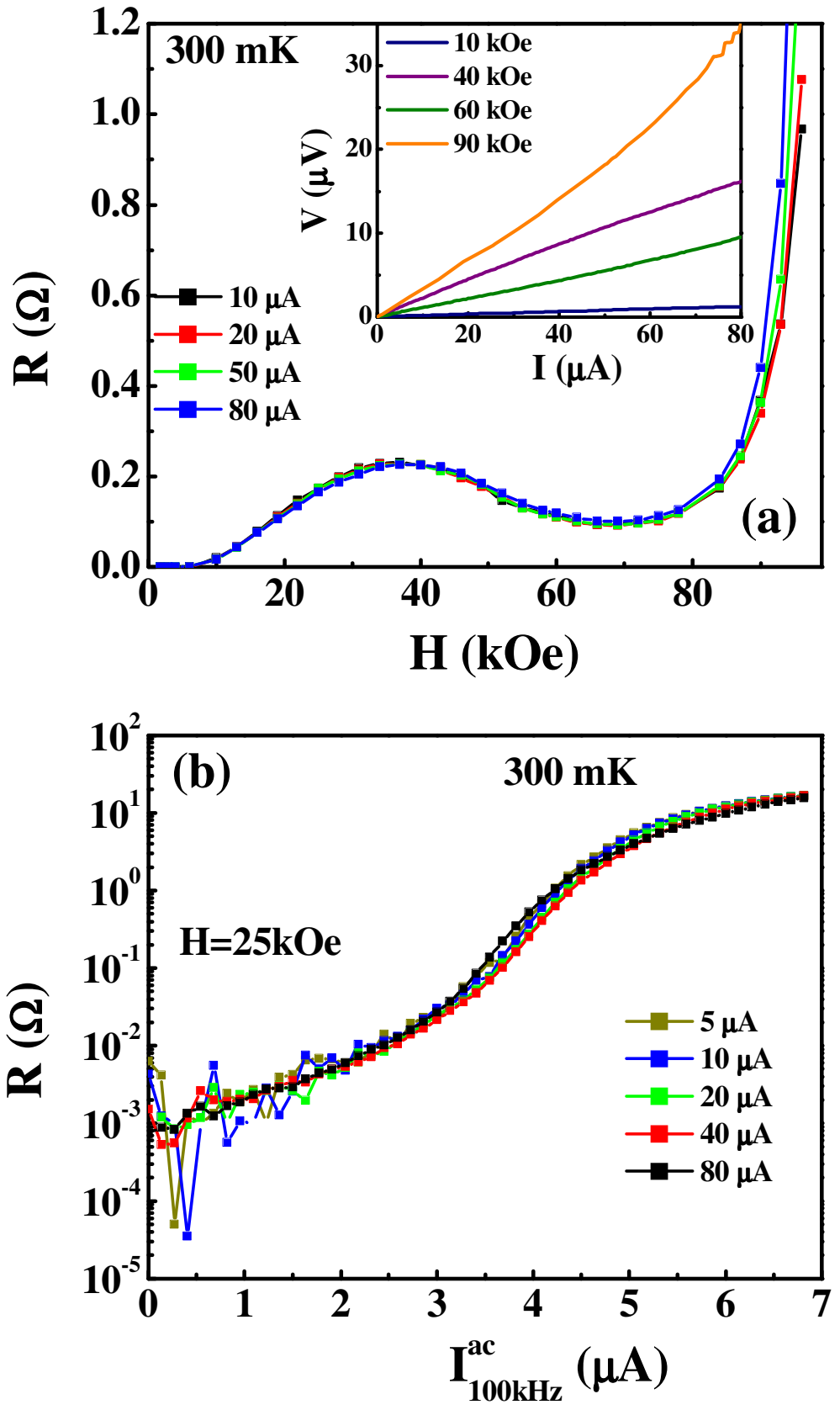

**Figure 9**